\documentclass[a4paper,11pt]{article}
\pdfoutput=1
\usepackage{amssymb,amsmath,amsfonts,makeidx,placeins,pbox,multirow}
\usepackage{graphicx,rotate,subcaption,color,slashed,cite,caption,epstopdf,verbatim}
\usepackage{longtable,tabu}
\usepackage{array}
\usepackage{pstricks}
\usepackage{color}
\usepackage{amsmath}
\usepackage{xfrac}
\usepackage{mathtools}
\usepackage{amssymb}
\usepackage{bbold}
\usepackage[normalem]{ulem}
\usepackage{caption}
\usepackage{subcaption}
\usepackage{dsfont}
\usepackage[b]{esvect}
\usepackage{tabularx}
\usepackage{array}
\usepackage[utf8]{inputenc}

\newcolumntype{P}[1]{>{\centering\arraybackslash}p{#1}}
\usepackage[colorlinks=true,
            linkcolor=magenta,
            urlcolor=blue,
            citecolor=blue]{hyperref}

\evensidemargin=\oddsidemargin
\newcommand{\beq}{\begin{equation}}
\newcommand{\eeq}{\end{equation}}
\newcommand{\bea}{\begin{eqnarray}}
\newcommand{\eea}{\end{eqnarray}}

\def\bar {\overline}

\def\to {\rightarrow}

\def\bea {\begin{eqnarray}}
\def\eea {\end{eqnarray}}
\def\n {\nonumber}

\def\barr{\begin{array}}
\def\earr{\end{array}}
\def\to{\end{rightarrow}}

\def\gev{\ensuremath{\mathrm{Ge\kern -0.1em V}}}

\usepackage{color}
\definecolor{amber}{rgb}{1,0.45,0}

\def\TeV{\ifmmode {\mathrm{\ Te\kern -0.1em V}}\else
                   \textrm{Te\kern -0.1em V}\fi}
\def\GeV{\ifmmode {\mathrm{\ Ge\kern -0.1em V}}\else
                   \textrm{Ge\kern -0.1em V}\fi}
\def\MeV{\ifmmode {\mathrm{\ Me\kern -0.1em V}}\else
                   \textrm{Me\kern -0.1em V}\fi}
\def\keV{\ifmmode {\mathrm{\ ke\kern -0.1em V}}\else
                   \textrm{ke\kern -0.1em V}\fi}
\def\eV{\ifmmode  {\mathrm{\ e\kern -0.1em V}}\else
                   \textrm{e\kern -0.1em V}\fi}
\let\tev=\TeV
\let\gev=\GeV

\def\ifb{\mbox{fb$^{-1}$}}
\usepackage{tikz}
\usetikzlibrary{positioning,arrows}
\usetikzlibrary{decorations.pathmorphing}
\usetikzlibrary{decorations.markings}

\newcommand{\fjtwo}  {\ensuremath{\mathrm{2L2F_j}}}
\newcommand{\fjthree}{\ensuremath{\mathrm{3L1F_j}}}
\newcommand{\pt}     {\ensuremath{p_{\mathrm{T}}}}
\newcommand{\Ht}     {\ensuremath{H_{\mathrm{T}}}}
\newcommand{\Lt}     {\ensuremath{L_{\mathrm{T}}}}
\newcommand{\St}     {\ensuremath{S_{\mathrm{T}}}}
\def\met{\ensuremath{p_{\mathrm{T}}^{\mathrm{miss}}}} 

\begin{document}

\tikzset{
vector/.style={decorate, decoration={snake,amplitude=.6mm}, draw=red},
scalar/.style={dashed, draw=blue},
fermion/.style={draw=black, postaction={decorate},
        decoration={markings,mark=at position .55 with {\arrow[draw=black]{>}}}}
}
\begin{center}
{\Large \bf Fatjet Signatures of Quintuplet Fermions at the LHC\\
\vspace{0.3cm}
} 
\vspace*{0.4cm} {\sf  $^{a}$Sourabh Dube \footnote{sdube@iiserpune.ac.in}, $^{b}$Nilanjana Kumar\footnote{nilanjana.kumar@gmail.com}, $^{a}$Shriyansh Ranjan\footnote{shriyanshranjan1@gmail.com} }\\
\vspace{6pt} {\small } {\em  $^{a}$Indian Institute of Science Education and Research (IISER), Pune, India \\$^{b}$Department of Physics, Chettinad Institute of Technology, Chettinad Academy of Research and Education, Tamilnadu 603103, India}\\
\normalsize
\end{center}
\vspace{-0.4cm}
\begin{abstract}
  This paper explores a simplified extension of the standard model featuring a neutral fermion quintuplet and a scalar quadruplet, which together generate neutrino masses through
  tree and loop level mechanisms. The quintuplet fermions decay into standard model gauge bosons via the scalars, producing unique collider signatures at the LHC characterized
  by multilepton and multijet final states. The study focuses on the pair production of quintuplet fermions in the $700–1200~\GeV$ mass range, where their decays produce highly
  boosted $W$ and $Z$ bosons identifiable as {\it fatjets}. Emphasis is placed on the production and decay of doubly charged fermions due to their higher cross section. Advanced
  jet substructure and kinematic techniques are applied to enhance sensitivity by reducing standard model backgrounds. 
  A detailed analysis of signal significance is performed in the two lepton, two fatjet (\fjtwo) and three lepton, one fatjet (\fjthree) channels for different masses of the
  fermion and the scalars, optimizing selection cuts to maximize signal efficiency over standard model backgrounds. The study found that both channels exhibit
  excellent performance, with significance exceeding $5\sigma$ under realistic conditions including a $50\%$ background uncertainty at integrated luminosities up to $3000~\ifb $.
\end{abstract}

\section{\label{sec:1}Introduction}
The standard model (SM) of particle physics, though highly successful, cannot explain several observed phenomena such as the small neutrino masses~\cite{Zyla:2020zbs},
the particle nature of dark matter, the muon anomalous magnetic moment~\cite{Muong-2:2021ojo}, and certain flavor anomalies~\cite{London:2021lfn} among many others.
Extensions of the SM introduce physics at or beyond the \TeV\ scale to address the shortcomings of the SM. Some popular examples are extensions of the SM with vector-like quarks and
leptons~\cite{Aguilar-Saavedra:2013qpa,Kumar:2015tna,Kearney:2012zi}, fermion multiplets~\cite{Foot:1988aq,Nomura:2017abu,Kumar:2019tat} or scalar
multiplets~\cite{Babu:2009aq,Chaudhuri:2013xoa,Du:2018eaw}. Beyond SM scenarios featuring both fermion and scalar multiplets, including cases with more than one set
of new multiplets, have been widely studied in the literature. The appearance of both fermion and scalar multiplets is natural in models of grand unified
theories~\cite{Georgi:1974sy}, Left-Right Symmetric scenarios~\cite{Senjanovic:1975rk,KumarAgarwalla:2018nrn},~Little Higgs \cite{Low:2002ws}, and Composite
Higgs Models~\cite{Vignaroli:2012sf}. Additionally, models such as the minimal R$\nu$MDM~\cite{Cai:2016jrl} incorporate a scalar or fermionic quintuplet and
a vector-like quadruplet fermion for unified explanations of neutrino mass and dark matter candidates.

In general, models with more than one multiplet or one type of BSM particle often allow interaction among themselves which results in distinct decay modes and thus the
collider signatures are unique. The mass difference between different BSM multiplets or particles plays a crucial role in the kinematics. These models have gained a lot
of interest recently. If we denote by $X$ and $Y$ the different BSM particles, then for example, in Ref~\cite{Dubey:2025sfh} $X$ and $Y$ are leptoquark and vector-like leptons.
In Ref~\cite{Ashanujjaman:2022cso} $X$ and $Y$ are two leptonic multiplets belonging to different $SU(2)_L$. In composite Higgs models~\cite{Kumar:2025rqs,Banerjee:2022xmu},
$X$ and $Y$ can be two different pNGB scalars or one pNGB scalar and one heavy resonance. In this paper, we examine a model with a neutral fermion quintuplet (with zero hypercharge)
and a scalar quadruplet; which allows for neutrino mass generation via tree-level and loop-level mechanisms~\cite{Ma:2014zda,Nomura:2017abu,Kumar:2019tat}. 

We choose a scenario where only one scalar quartet ($\phi_4$) and one quintuplet fermion ($\Sigma_R$) is present. This model provides a unique signature at the LHC
and future collider experiments~\cite{Kumar:2021umc} in multilepton and multijet final states. Once produced at the LHC, these quintuplet fermions decay to SM gauge
bosons via the scalars where the fermions are heavier than the scalars. It was shown in Ref~\cite{Kumar:2019tat} that the final states with lepton and jets suffer
from very small cross sections, especially when the mass of the fermion quintuplet is around 1~\tev.

Current searches for the exotic fermions and scalars have put strong bounds on their masses. For example, the ATLAS experiment sets a lower limit on fermion mass
of 910~\gev~\cite{ATLAS:2022yhd}. The CMS experiment sets constraints on the fermion masses for varied coupling scenarios, with the lower limit being as high as
1065~\gev~\cite{CMS:2022nty} in specific cases. The direct search limit on charged Higgs bosons of $80~\gev$ comes
from the LEP experiments~\cite{ALEPH:2013htx}. Both ATLAS and CMS have conducted extensive searches targeting multiple production and decay modes of the scalars. 
A lower limit of $\sim 200~\gev$ is placed on the masses of charged scalars from various channels, when the scalars decay to the SM gauge bosons~\cite{ATLAS:2022zuc,CMS:2021wlt,ATLAS:2023sua}. 
Limits exist on the production cross section of the charged scalars when they decay to SM fermions~\cite{ATLAS:2022pbd,CMS:2019bfg}. These searches have placed a
stricter lower limit on the charged Higgs mass of $\sim 400~\gev$. 
The LHC limits are not directly applicable on the model under study since the quintuplet fermions decay through scalars.
However, to analyze the fatjet signatures we consider the mass of the fermion quintuplet $\geq 700~\gev$, and the mass of the scalar multiplet at $\geq 600~\gev$,
which is above the current lower bounds.

Standard searches for fermion multiplets are proposed widely in the 
literature~\cite{Das:2020uer,Yu:2015pwa,Chen:2013xpa,Ozansoy:2019kdq,Guo:2016hjt,Zeng:2016tmw}.
In this paper we investigate the pair production of the quintuplet fermions where the fermions decay via the charged scalars to the SM gauge bosons.
If the mass of the fermion multiplet is $\geq 700$ \GeV, then the $W/Z$ bosons from the decay of the scalars are expected to be highly 
boosted and they can be identified as {\it fatjets}. Fatjets are the resultant objects of  jet clustering algorithms using a large radius parameter,
thus making them {\it fat}. Fatjet searches at the LHC have become a critical tool for probing BSM particles. Fatjet signatures are studied in 
various BSM models with fermion multiplets to achieve sensitivity to masses up to and beyond \TeV\ scale~\cite{Choudhury:2021nib,Ashanujjaman:2021zrh,Dey:2022tbp}.
Here we primarily focus on the production of the doubly charged fermions because of its higher production cross section as compared to the other pair production
modes. We show that advanced jet reconstruction and jet substructure~\cite{Kumar:2022lqp} techniques help to suppress SM backgrounds and enhance the sensitivity
of LHC experiments.

\section{\label{sec:2}The Model}

We consider a simplified model where the SM is extended with a quintuplet fermion ($\Sigma_R$) and a quartet scalar ($\phi_4$). 
This model has been studied previously in detail \cite{Nomura:2017abu,Kumar:2019tat,Kumar:2021umc}. 
The quintuplet fermion $\Sigma_R$ of hypercharge $0$ can be expressed as,
$\Sigma_R \equiv \left[\Sigma_1^{++},\Sigma_1^{+},{\Sigma^0}, \Sigma_{2}^-, \Sigma_{2}^{--} \right]_R^T$
and the scalar multiplets with hypercharge $\sfrac{1}{2}$ is given by,
$\Phi_4 = \left( \varphi^{++}, \varphi^{+}_2, \varphi^{0}, \varphi^{-}_1 \right)^T$.
The gauge coupling of $\Sigma_R$ is obtained as,
\begin{eqnarray}
\mathcal{L}_{gauge,\Sigma} &\supset& (e A_\mu + g c_w Z_\mu) ~\bigg[ 2 (\bar{\Sigma}^{++}_R \gamma^\mu \Sigma^{++}_R + \bar{\Sigma}^{++}_L \gamma^\mu \Sigma^{++}_L) +\bar{\Sigma}^{+}_R \gamma^\mu \Sigma^{+}_R + \bar{\Sigma}^{+}_L \gamma^\mu \Sigma^{+}_L \bigg] - \\
&& g ~\bigg\{ \bigg[ \sqrt{2} (\bar{\Sigma}^{+}_R \gamma^\mu \Sigma^{++}_R - \bar{\Sigma}^{+}_L \gamma^\mu \Sigma^{++}_L)+ \sqrt{3} (\bar{\Sigma}^{0}_R \gamma^\mu \Sigma^{+}_R - \bar{(\Sigma^{0}_R)^C} \gamma^\mu \Sigma^{+}_L) \bigg] W^-_\mu + h.c. \bigg\}
\label{eq:gauge_sigma}
\end{eqnarray}
The interaction of the quartet scalar with the gauge bosons is obtained from the kinetic term as
\begin{eqnarray}
\mathcal{L}_{gauge,\Sigma} &\supset& \frac{\sqrt{3}}{2} \frac{g^2}{cw} v_4 \left(2 - s_W^2 \right)W^+_\mu Z^\mu \phi_1^- + \frac{g^2}{cw} v_4 s_W^2 W^+_\mu Z^\mu \phi_2^- \\
&&+ \sqrt{\frac{3}{2}} {g^2} v_4 W^+_\mu W^{+\mu} \phi^{--} + ~h.c.
\end{eqnarray}
The scalar also couples with the quintuplet fermion via the Yukawa term.
\begin{eqnarray}
-{\cal L}_{yuk} \supset (y_{\nu})_{ij} [\bar L_{L_i} \tilde\Phi_4 \Sigma_{R_j} ]+{\rm h.c.}
\end{eqnarray}
Upon expansion, 
\begin{eqnarray}
-{\cal L}_{yuk}&=&(y_\nu)_{ij} \left[\bar\nu_{L_i}\left(\Sigma^{++}_{1R_j}\varphi^{--}+\frac{\sqrt3}{2}\Sigma^+_{1R_j}\varphi_1^- +\frac{1}{\sqrt2}\Sigma^0_{R_j}\varphi^{0*}+\frac{1}{2}\Sigma^-_{2R_j} \varphi_{2}^+ \right)\right.\\
&& \left. +
\bar\ell_{L_i} \left( \frac12 \Sigma^+_{1R_j} \varphi^{--} + \frac{1}{\sqrt2}\Sigma^0_{R_j} \varphi_1^- + \frac{\sqrt3}{2}\Sigma^-_{2R_j} \varphi^{0*} + \Sigma^{--} _{2R_j}\varphi_{2}^{+}  \right)\right] +{\rm h.c.}
\label{eq:yukawa}
\end{eqnarray}
Here, we have neglected the mixing among the individual components of the multiplets and we choose the components of
the quintuplet fermion as well as the scalar multiplets to be degenerate, which we denote as $m_\Sigma$ and $m_\phi$, respectively\cite{Nomura:2017abu}. We also consider 
$m_\Sigma>m_\phi$, which is naturally implied form the muon anomalous magnetic moment 
measurement, as shown in \cite{Kumar:2019tat}. For the complete Lagrangian 
we refer to  Ref.~\cite{Nomura:2017abu}.
 
The charged fermions $\Sigma^{\pm}$ and $\Sigma^{\pm\pm}$ decay via the following modes:
\bea
\Sigma^{\pm} &\rightarrow &\phi_2 (\phi_1)^{\pm} \nu (\bar{\nu}), \phi^{\pm \pm} l^{\mp}, \phi^{0} l^{\pm} \\
\Sigma^{\pm \pm} &{\rightarrow}&\phi^{\pm \pm} \nu (\bar{\nu}), \phi_2^{\pm} l^{\pm}
\label{eqn:decayss2}
\eea
where $l=e,\mu,\tau$.
The branching ratios are $\Sigma^{\pm \pm} \rightarrow \phi^{\pm \pm} \nu (\bar{\nu})= \Sigma^{\pm \pm} \rightarrow \phi_2^{\pm} l^{\pm}=50\%$.
The scalars further decay to SM gauge bosons with 100\% branching ratio:
\bea
\phi_2^{\pm}(\phi_1^{\pm})&\rightarrow & W^{\pm} Z \n \\
\phi^{\pm \pm}~~~~ &\rightarrow & W^{\pm} W^{\pm} \n \\
\phi^{0}~~~~~~ &\rightarrow & W^{+} W^{-}.
\label{eqn:phidecay}
\eea
\section{\label{sec:3}Signal and Background at the LHC}
\begin{figure}[h]
  \centering
  \begin{subfigure}[b]{0.48\linewidth}
    \includegraphics[height=4.5cm,width=\linewidth]{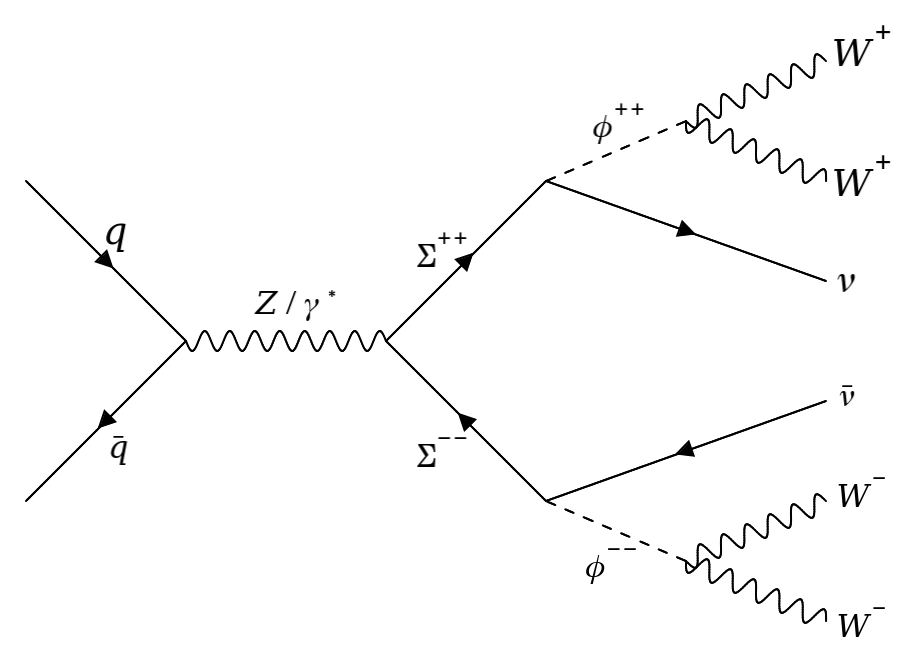}
    \caption{Quark Initial State}
  \end{subfigure}
  \begin{subfigure}[b]{0.42\linewidth}
    \includegraphics[height=4.5cm,width=\linewidth]{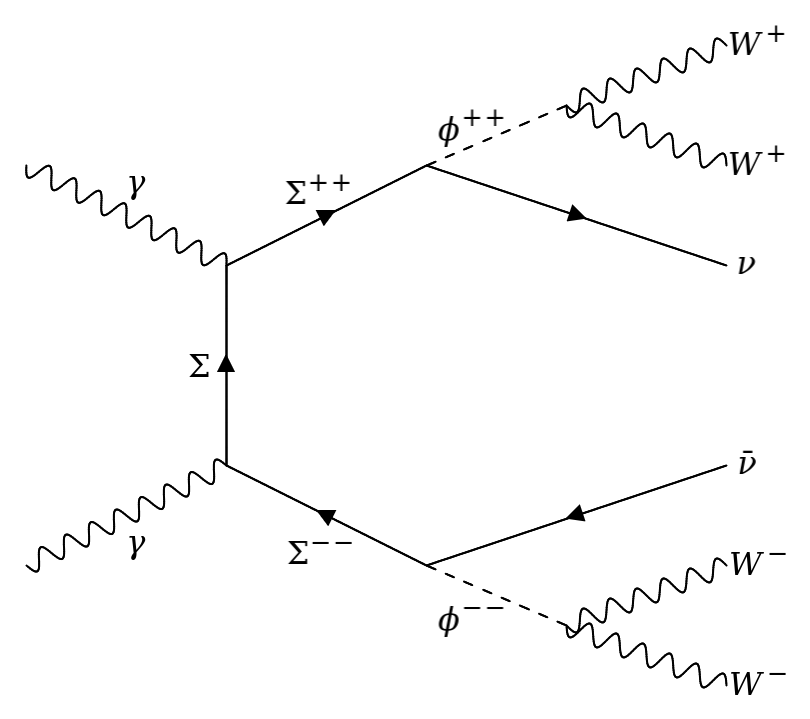}
    \caption{Photon Initial State}
  \end{subfigure}
  \caption[Feynman diagram for the quintuplet model]{Feynman diagrams of the doubly charged fermion pair production at the LHC, where $q$ stands for $u,d,c,s,b$ quarks.}
  \label{fd_quint}
\end{figure}

The charged and neutral scalars can be pair produced at the LHC via photon and $Z$ boson mediated
Drell–Yan (DY) processes and also via a $t$-channel process. 
The cross section of $pp\rightarrow\Sigma^{\pm}\Sigma^{\pm}$ is significantly smaller than $pp\rightarrow\Sigma^{\pm\pm}\Sigma^{\mp\mp}$ because 
the production cross section varies with the fourth power of the charge.
A detailed discussion of the pair production cross section is given in Ref.~\cite{Kumar:2021umc}.
Here we investigate the pair production of $\Sigma^{\pm\pm}\Sigma^{\mp\mp}$ at $pp$ collisions at the centre-of-mass energy of 14~\tev.
The photon parton distribution function (PDF) for the proton is included to the initial state of the process given its significant contribution to
the production cross section. The diagrams leading to the pair production of $\Sigma^{\pm\pm}\;\Sigma^{\mp\mp}$ are shown in Fig.~\ref{fd_quint} following
the decays (one among many) of the particles. 

The inclusion of the photon PDF~\cite{Manohar:2024ndc} raises the question of the choice of PDF to use when generating simulation samples. 
To investigate this, we used different versions of NNPDF23 \cite{Carrazza:2013bra,Ball:2013hta}. We compare a quark-only initial state,
a photon-only initial state, and an inclusive initial state using three different LO and NLO PDFs for two different masses of $\Sigma^{\pm\pm}$,
$400\GeV$ and $1000\GeV$. For this, \textsc{MadGraph}~\cite{Alwall:2014hca,Frederix:2018nkq} incorporated with \textsc{lhapdf}~\cite{Buckley2015}
is used. We observe a significant increase in the production cross section by using a photon PDF for $m_{\Sigma}=1000\GeV$, and modest increase at $400\gev$
irrespective of the choice of PDFs. We use the cross section calculated by using NNPDF2.3QED at NLO, specifically NNPDF23\_nlo\_as\_0119\_qed, and show
this in Table~\ref{quint_pdf}. The increase in the production cross section after including the photon PDF is also shown and the uncertainties in the
cross sections are found to be $<1\%$.

\begin{table}[h]
  \centering
  \renewcommand{\arraystretch}{1.5}
  \begin{tabular}{|c|c|c|c|c|c|}

    \hline
    \multirow{2}{1.4cm}{$m_{\Sigma^{\pm\pm}}$ (\GeV)} & \multirow{2}{2.1cm}{PDF used} & \multicolumn{3}{|c|}{Production cross section ($pb$)} & \multirow{2}{2.7cm}{\small{Relative increase in cross section (\%)}} \\
    \cline{3-5}
    & & \small{quark-only} & \small{photon-only} & \small{inclusive} & \\
    \hline
    $400$ & \footnotesize{NNPDF23\_nlo\_as\_0119\_qed} & $0.1851$ & $0.0094$ & $0.1944$ & $5.0$ \\ 
    \hline
    $1000$ & \footnotesize{NNPDF23\_nlo\_as\_0119\_qed} & $0.001767$ & $0.000327$ & $0.002089$ & $18.2$ \\ 
    \hline
  \end{tabular}
  \caption{Pair production cross section of $\Sigma^{++}\Sigma^{--}$ is compared for initial states that are quark-only, photon-only, and inclusive in 14~\tev\ $pp$ collisions.
    The percentage increase in the inclusive case is calculated with respect to the quark-only initial state.
  }
  \label{quint_pdf}
\end{table}

The different decays of the quintuplet fermions and the quadruplet scalars give rise to the following final states involving bosons and leptons.
\begin{align}
p p \rightarrow &\;\Sigma^{\pm\pm}\Sigma^{\mp\mp}\rightarrow \phi^{++} \nu  \phi^{--} \bar{\nu} \rightarrow W^+ W^+ W^- W^- + \nu\bar{\nu} \\
p p \rightarrow &\;\Sigma^{\pm\pm}\Sigma^{\mp\mp}\rightarrow \phi_2^{+} l^+  \phi_2^{-} l^- \rightarrow W^+ Z W^- Z + l^+ l^- \\
p p \rightarrow &\;\Sigma^{\pm\pm}\Sigma^{\mp\mp}\rightarrow \phi^{++} \nu  \phi_2^{-} l^- \rightarrow W^+ W^- W^- Z + l^- + \nu
\end{align}

Three benchmark points for the signal are considered for the analysis:
\begin{itemize}
\item {\bf Low-mass}: $m_{\Sigma^{++}}=700\GeV$, $m_{\phi^{++}}=600\GeV$
\item {\bf Med-mass}: $m_{\Sigma^{++}}=1200\GeV$, $m_{\phi^{++}}=600\GeV$
\item {\bf High-mass}: $m_{\Sigma^{++}}=1200\GeV$, $m_{\phi^{++}}=1100\GeV$
  \end{itemize}
The $W$ and $Z$ bosons in the final state are often boosted when they decay hadronically, and thus can be identified as fatjets.
The first channel can provide up to four $W$-fatjets whereas the second and third channel generate up to four fatjets which can be either $W$- or
$Z$-fatjets. Although the expected production yield is identical for each channel, we opt for channel 1 -- characterized by four $W$ bosons.
Having a single type of fatjet will streamline the analysis. Figure~\ref{fd_quint} shows the pair production process of the doubly charged fermion
leading to four $W$'s in the final state. 

The efficiency of reconstructing and identifying the $W$-fatjets increases when the mass of the parent particle is $\sim 1\TeV$ since the $W$ is boosted and
the decay products of the $W$ become increasingly collimated. At a hadron collider, the presence of a lepton in the final state allows a large reduction
in the SM background. We combine the requirement of a $W$-fatjet and a lepton to maximize our advantage and choose to study the following two final states:
\begin{enumerate}
    \item $\fjtwo$: Two leptons (L$=e,\mu,\tau$) with two $W$-fatjets 
    \item $\fjthree$: Three leptons (L$=e,\mu,\tau$) with one $W$-fatjet
\end{enumerate}
These final states ensure that we reduce the SM background significantly, while at the same time maintaining good efficiency for lower $\phi$ masses.
Table~\ref{xs_sig} shows the cross section ($\sigma$) $\times$ branching ratio ($\mathcal{B}$) for the three chosen benchmark points. We have assumed
$\mathcal{B}(\Sigma^{++}\rightarrow\phi^{++} \nu)=100\%$.

\begin{table}[h!]
  \centering
  \renewcommand{\arraystretch}{1.5}
  \begin{tabular}
{|c|c|c|c|}
\hline
    \fjtwo & $m_{\Sigma^{++}},m_{\phi^{++}}$ (\GeV) & $\sigma\times\mathcal{B}$ (pb)\\
    \hline
    \multirow{3}{2.1cm}{~}& 1200,1100 (high-mass)& $1.455\times10^{-4}$ \\
    \cline{2-3}
    & 1200,600 (med-mass) & $1.455\times10^{-4}$\\
    \cline{2-3}
    & 700,600 (low-mass) & $3.627\times10^{-3}$\\
    \hline
\fjthree & $m_{\Sigma^{++}},m_{\phi^{++}}$ (\GeV) & $\sigma\times\mathcal{B}$ (pb)\\
\hline
    \multirow{3}{2.1cm}{~} & 1200,1100 (high-mass)& $4.405\times10^{-5}$\\
    \cline{2-3}
    & 1200,600 (med-mass)& $4.405\times10^{-5}$\\
    \cline{2-3}
    & 700,600 (low-mass) & $1.098\times10^{-3}$\\
    \hline
  \end{tabular}
  \caption{The cross section$\times $branching ratio in the channels of interest for the three benchmark points.}
  \label{xs_sig}
\end{table}

Several SM processes can give rise to backgrounds in the \fjtwo\ final state. The dominant contribution comes
from $t\bar{t}Z$~\cite{CMS:2019nos} where the $Z$ produces two leptons, and two $W$ bosons along with $b$ quarks originate from the top decay.
Similarly, the $ZVV$~\cite{PhysRevLett.125.151802} process, where $V$ denotes a $W$ or a $Z$, also contributes to the background. Due to the
large cross section, $t\bar{t}$~\cite{PhysRevD.104.092013} production with additional jets is also a source of background.
This process contributes when the $W$ bosons decay leptonically, and the jet arising from the $b$-quark is misidentified as a fatjet.

The possible backgrounds for the $\fjthree$ final state are: $WZ+$jets~\cite{ATLAS:2021jgw,CMS:2024ilq}, $t\bar{t}Z$~\cite{CMS:2019nos},
$t\bar{t}W$~\cite{CMS:2022tkv} and $t\bar{t}$+jets. The $WZ$+jets contributes the most when both of the bosons decay leptonically producing three leptons
and the one of the jets is misidentified as a fatjet. In $t\bar{t}W$ process, the three leptons can arise from any of the $W$ bosons, or from the
semi-leptonic decays of $b$-hadrons, and any of the $W$'s can give rise to the fatjet.
The requirement of three leptons makes the contribution from $t\bar{t}$ + jets negligible.

\section{\label{sec:4}Analysis of Fatjet Signals at the LHC}

Simulation for the signal and background processes at the LHC was done with \textsc{MadGraph5\_amc@nlo}~\cite{Alwall:2014hca} with
\textsc{Pythia \;8.3}~\cite{10.21468/SciPostPhysCodeb.8, 10.21468/SciPostPhysCodeb.8-r8.3} used for hadronization of the event. The
anti-$k_t$ jet clustering algorithm~\cite{Cacciari:2008gp} as implemented in the \textsc{FastJet}~\cite{Cacciari:2011ma} package within
the \textsc{Delphes}~\cite{deFavereau2014} framework is used for clustering jets and fatjets. For the signal, $5\times 10^4$ collision events
are simulated for each benchmark point. For each of the $t\bar{t}Z$, $t\bar{t}W$, and $WZ+$jets processes we generate $5\times 10^5$ events.
For $t\bar{t}$+jets and $ZVV$ processes, $1\times 10^6$ and $2\times 10^6$ events are produced respectively.

The initial object selections are as follows:
\begin{itemize}
\item Hadronically-decaying $\tau$ leptons ($\tau_h$) are reconstructed from the visible daughter particles (i.e. excluding the neutrino). The visible
  daughter particles thus exclude electrons and muons. Only $\tau_h$ originating from a signal particle or a gauge boson are considered.
\item  Electrons and muons are also required to originate either from a signal particle or a gauge boson, or from a $\tau$ lepton which
  has originated from a signal particle or a gauge boson.
\item We require $\pt(e,\mu)>10\GeV$, $|\eta(e,\mu)| <2.4$ $\pt(\tau_h)>20\GeV$, $|\eta(\tau_h)|<2.3$. The leading lepton in the event is required
  to satisfy $\pt>30\gev$ to mimic a trigger requirement.
\item Regular jets are clustered with the jet radius parameter $R=0.5$ and must satisfy $\pt>30\gev$ and $|\eta|<2.4$.
\item Fatjets are clustered with $R=0.8$ and must satisfy $\pt>200\gev$ and $|\eta|<2.4$.
\item Regular and fatjets are required to be away from the selected leptons by imposing $\Delta R(lj) > 0.4$,
  where $\Delta R = \sqrt{(\Delta\eta_{lj})^2+(\Delta\phi_{lj})^2}$.
\end{itemize}
\subsection{Kinematic Variables}
The signal and background events can be discriminated by using additional kinematic variables. These are defined below:

\begin{itemize}
\item $\met$: defined as the magnitude of the vector sum of transverse momenta of all the observed particles
  in the event --  $\met = \left| \sum\limits_{i=\mathrm{observed}} \vv{\pt}^i \right|$
\item $\Lt$: defined as the scalar sum of $\pt$ of all the charged leptons that constitute the final state --
  $\Lt = \sum\limits_{i=\mathrm{leptons}} \pt^i $
  \item $\Ht$: defined as the scalar sum of $\pt$ all the regular jets -- $\Ht = \sum\limits_{i=\mathrm{jets}} \pt^i$
   \item $\St$: defined as the sum of $\Lt$, $\Ht$ and $\met$ -- $\St = \Lt + \Ht + \met$
  \item $m_{min}$: defined as the minimum of all of the invariant masses of lepton pairs, irrespective of their charge or flavor.
  \item $dR_{min}$: defined as the minimum of all of the angular distances between lepton pairs in the event.
  \item Jet momentum fraction ($JMF$): defined as the ratio of the scalar sum of \pt\ of the fatjets to the \Ht\ --
\begin{equation*}JMF = \frac{\sum\limits_{i=\mathrm{fatjets}}\pt^i}{\Ht} \end{equation*}
    
\end{itemize}
Aside from these quantities, the analysis uses several jet substructure variables to select the desired fatjets.

\begin{figure}[h!]
  \centering
    \includegraphics[width=0.49\linewidth]{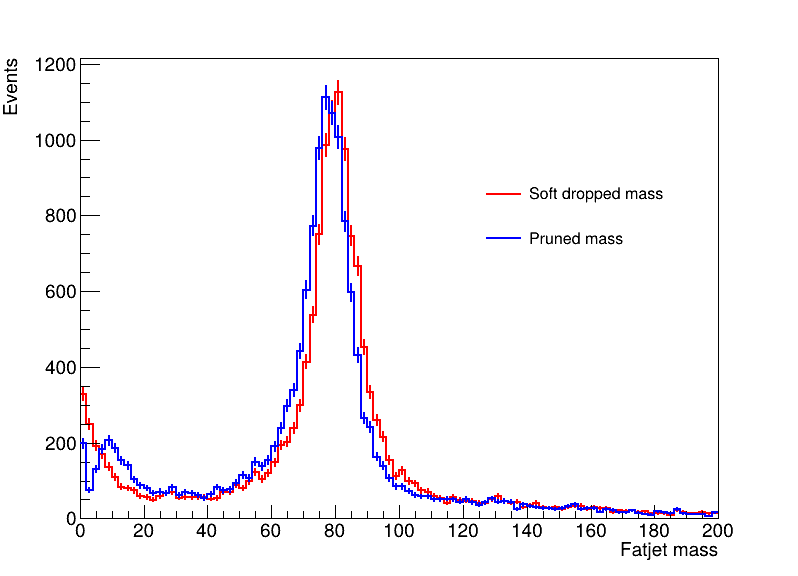}
    \includegraphics[width=0.49\linewidth]{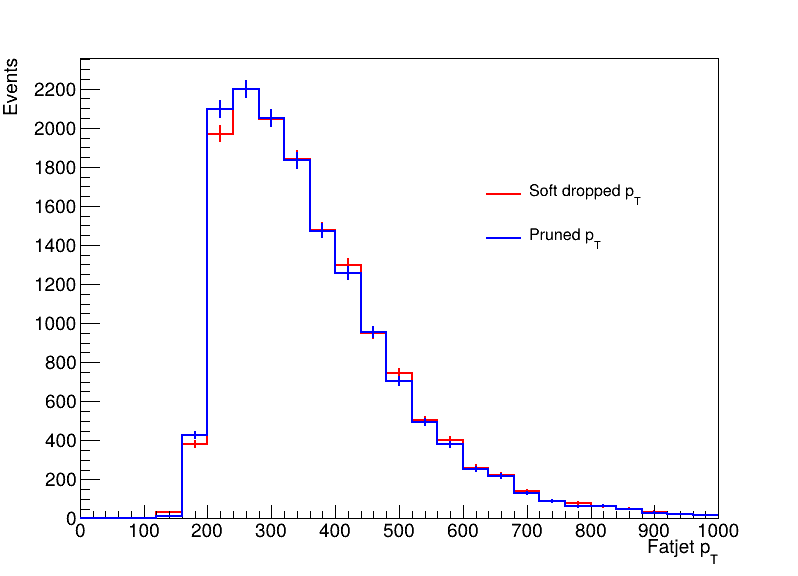}
  \caption{ Jet mass (left) and \pt\ (right) of the leading fatjet for two different jet grooming algorithms: soft drop and jet pruning.
    The low-mass benchmark point is used here. }
  \label{quint_ovl_groom}
\end{figure}

\paragraph{Jet Grooming:}
Since fatjets are considerably wider than regular jets, radiation from the underlying event or pileup can cause
an overestimation of the fatjet invariant mass and degrade the mass resolution. Jet grooming algorithms are
applied to the fatjets to account for these effects by removing unwanted soft radiation. There are two algorithms of jet grooming
that we consider: the soft drop algorithm~\cite{Larkoski2014} and the jet pruning algorithm~\cite{PhysRevD.81.094023}.
To compare these two, the resulting fatjet mass and $\pt$ after grooming is examined for the low-mass benchmark point and shown
in Fig.~\ref{quint_ovl_groom} (left). We find that the performance of the two algorithms is very similar. We choose the soft drop
algorithm, as the peak of the jet mass distribution is closer to $m_W$, and the jet \pt\ is slightly higher.

\begin{figure}[h!]
  \centering
  \includegraphics[width=0.49\linewidth]{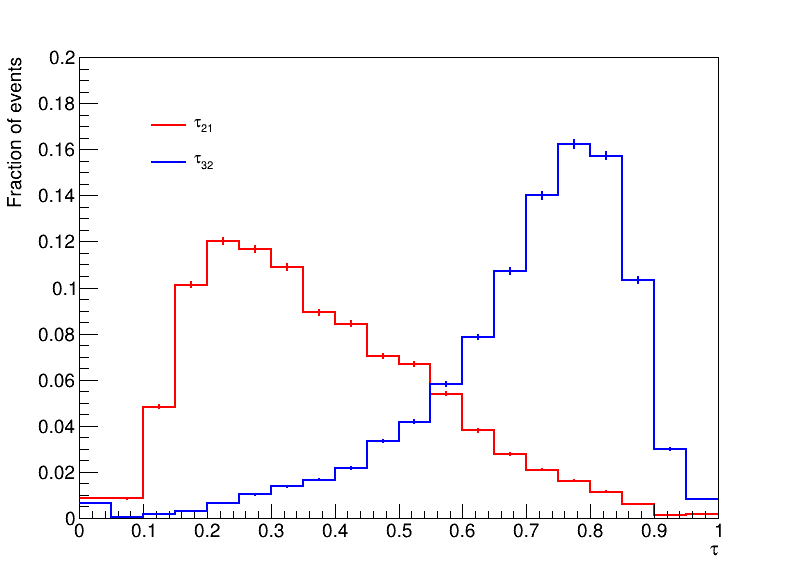}
  \includegraphics[width = 0.49\linewidth]{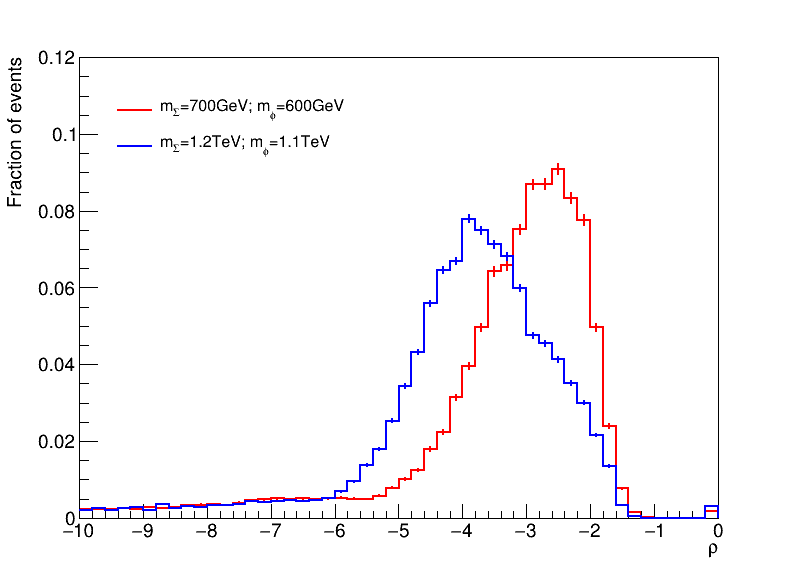}
  \caption{ The $\tau_{21}$ and $\tau_{32}$ variables (left) for the leading fatjets in the low-mass benchmark point and 
    the relative mass $\rho$ (right) for the leading fatjets in the low-mass and high-mass benchmark points.}
    \label{ovl_sb}
\end{figure}

\paragraph{n-subjettiness:}
The n-subjettiness variables ($\tau_n$)~\cite{Napoletano2018} can be used by taking ratios of subsequent n-subjettiness
($\tau_{p(p-1)} = \tau_{p} / \tau_{p-1}$) to obtain a relative probability of the jet being $p$-pronged or ($p-1$)-pronged.
Figure~\ref{ovl_sb} (left) shows the behavior of $\tau_{21}$ and $\tau_{32}$ for the low-mass and high-mass benchmark points.
The low value of $\tau_{21}$ and high value $\tau_{32}$ show that the $W$-fatjet has a $2$-prong nature. 

\paragraph{Relative Mass}
The relative mass ($\rho$) of a fatjet is calculated using the the soft drop mass ($m_{SD}$) and jet \pt\ as $\rho = 2 \ln(m_{SD}/\pt)$. This helps avoid
regions where non-perturbative effects lead to discrepancies in generator predictions~\cite{PhysRevD.101.052007}. 
Figure~\ref{ovl_sb} (right) shows the behavior of the $\rho$ variable of the leading fatjet for the low-mass and high-mass benchmark points.
We see that the non-perturbative effects are less significant for the high-mass point.

\subsection{$\fjtwo$ Channel}
\begin{figure}[h]
\includegraphics[width=0.33\linewidth]{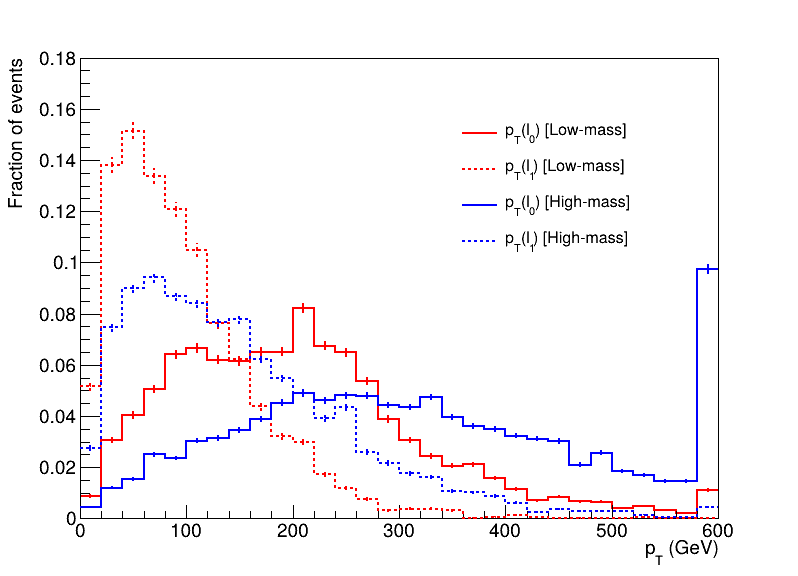}
\includegraphics[width=0.33\linewidth]{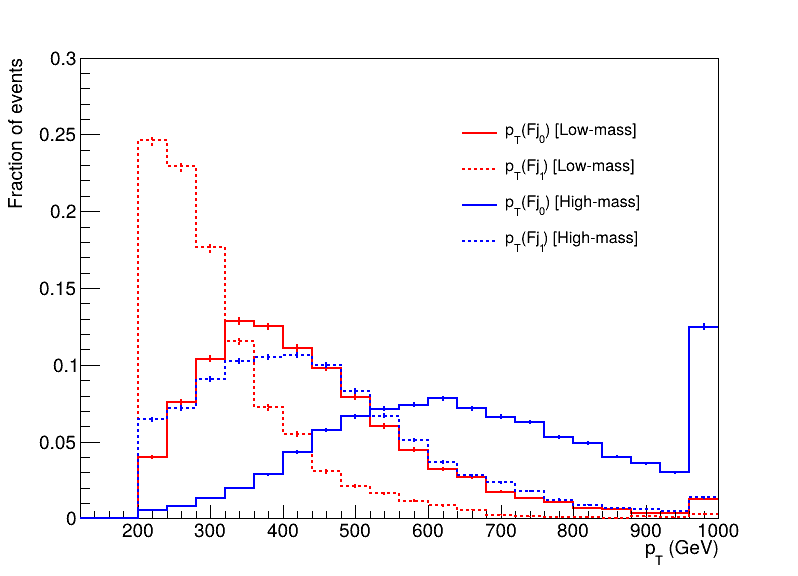}
\includegraphics[width=0.33\linewidth]{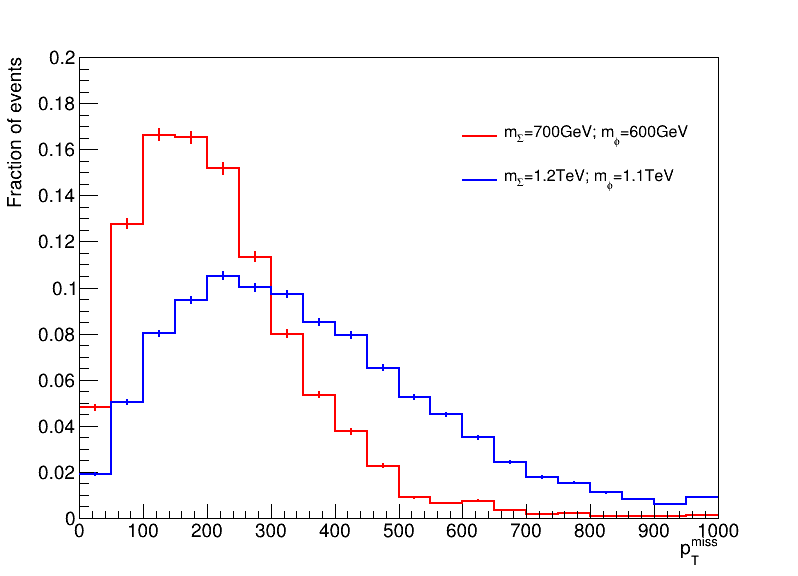}
\caption{The $\pt$ of the leading and subleading lepton (left), \pt\ of the leading and subleading fatjet (middle), and $\met$ (right) are shown for
  the low-mass and high-mass benchmark points. The distributions in each figure are normalized to have unit area.}
\label{fig:dist2}
\end{figure}

The \fjtwo\ final state arises from the process $pp\rightarrow\Sigma^{++}\Sigma^{--}\rightarrow\phi^{++}\nu\phi^{--}\bar{\nu}\rightarrow W^+W^+\nu W^-W^-\bar{\nu}$
when two of the four $W$ bosons decay leptonically and the other two decay hadronically.
In Fig.~\ref{fig:dist2} we show the distribution of $\pt$ of the leading and subleading lepton and fatjet, and the \met\ in signal events. As expected,
the high-mass signal benchmark point shows larger values of each of these variables.

\begin{figure}[h!]
\includegraphics[width=0.49\linewidth]{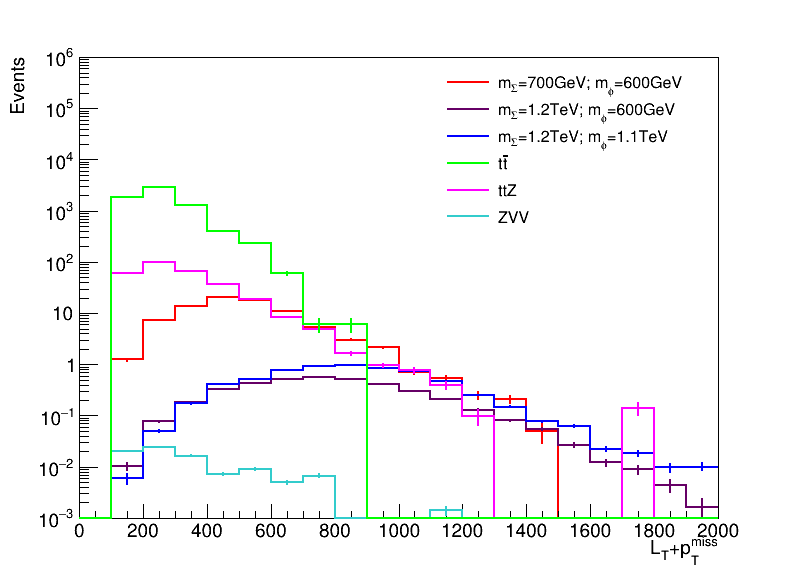}
\includegraphics[width=0.49\linewidth]{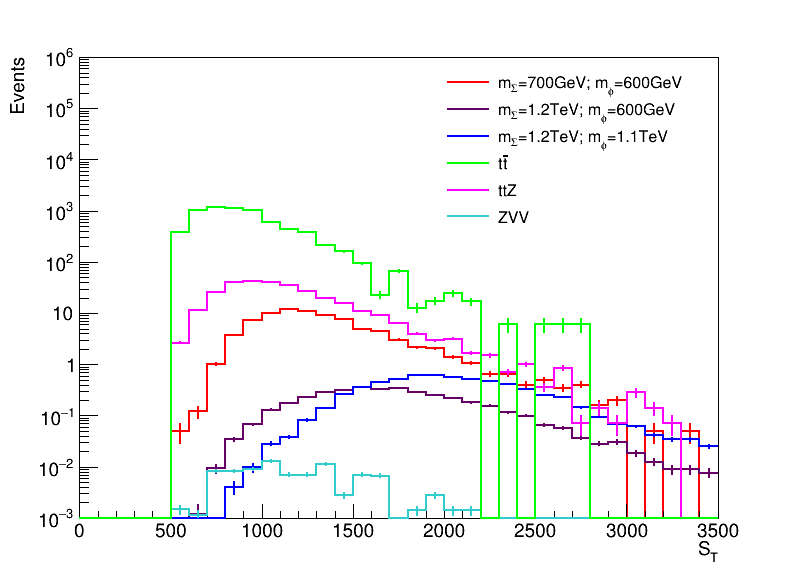}
\includegraphics[width=0.49\linewidth]{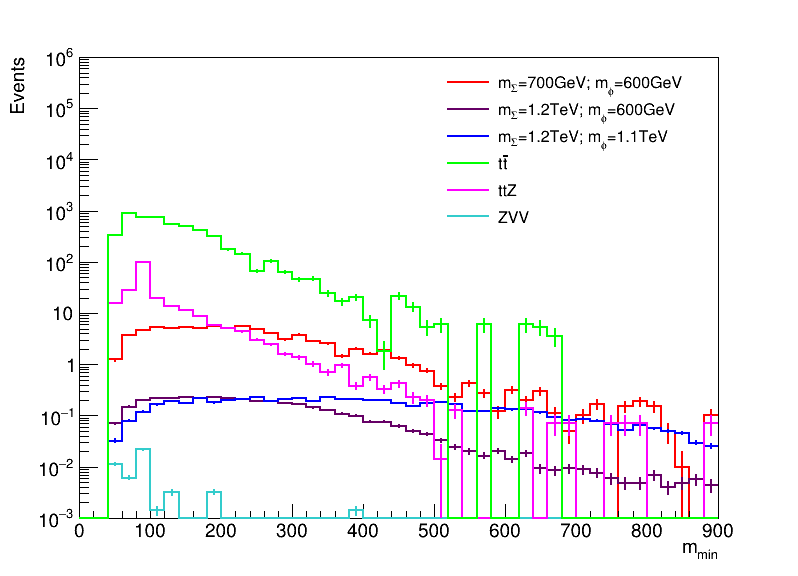}
\includegraphics[width=0.49\linewidth]{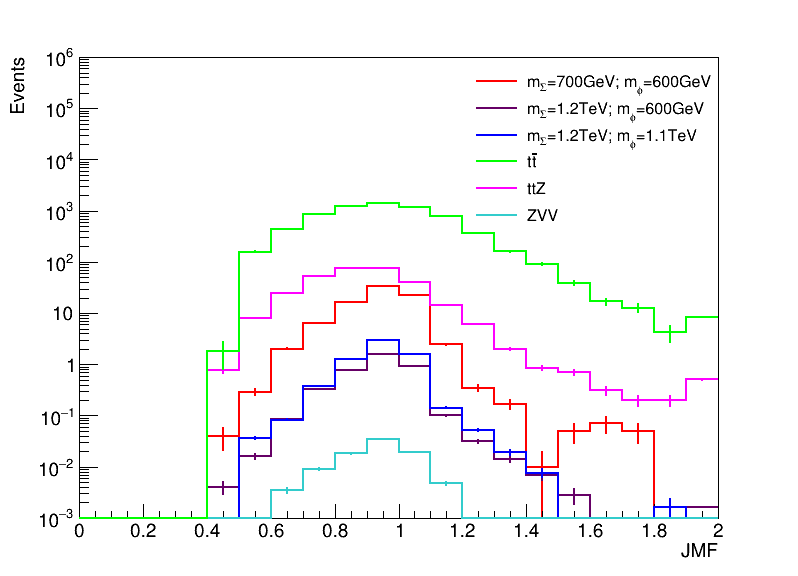}

\caption{The distributions of $\Lt+\met,\St$ (top row) and $m_{min}, JMF$ (bottom row) are shown for signal and background processes after
  the S1 selections are imposed in the \fjtwo\ final state.}
\label{fig:dist3}
\end{figure}

\begin{figure}[h!]
  \includegraphics[width=0.33\linewidth]{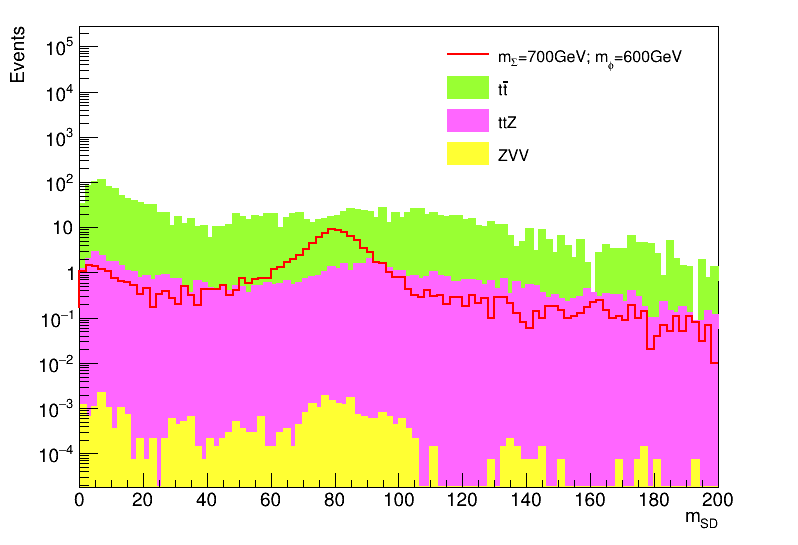}
  \includegraphics[width=0.33\linewidth]{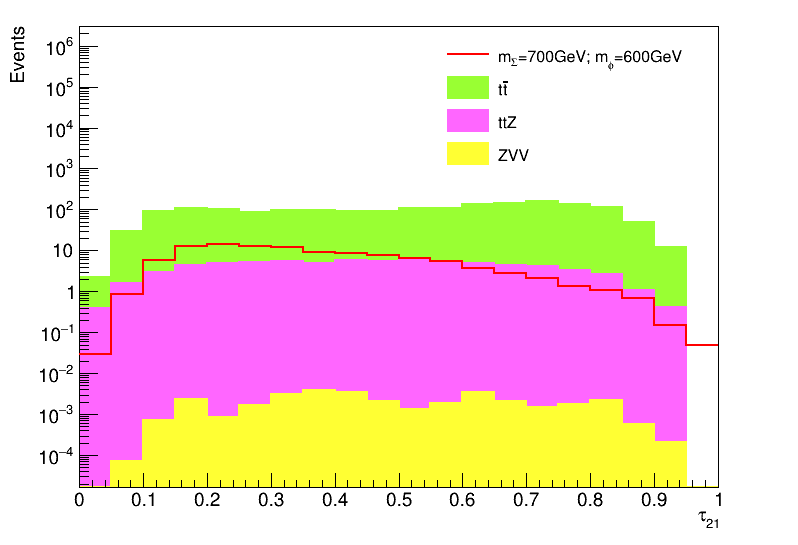}
  \includegraphics[width=0.33\linewidth]{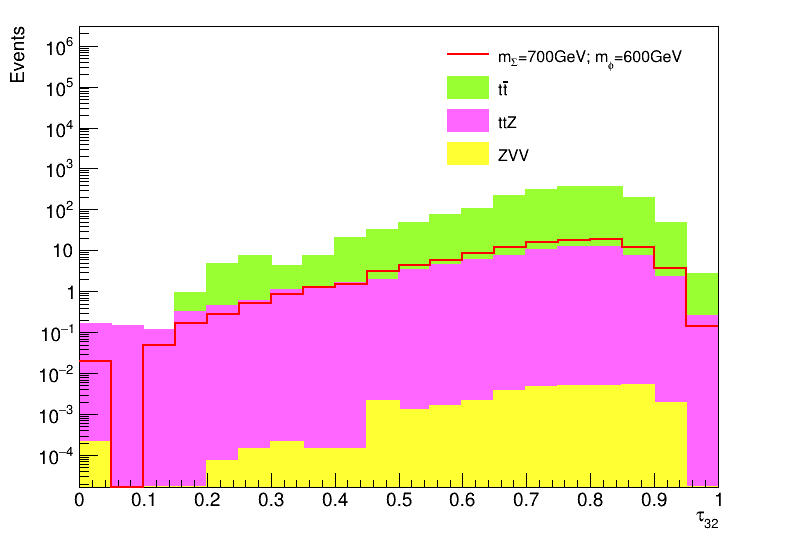}    
  \includegraphics[width=0.33\linewidth]{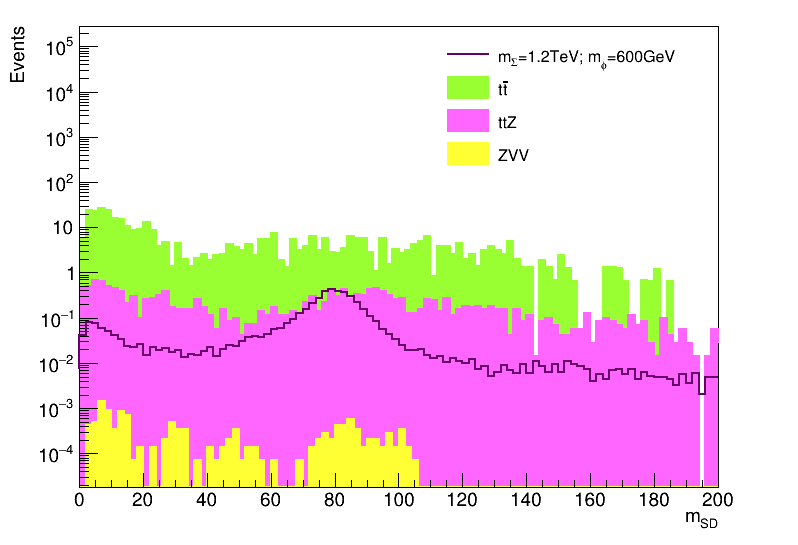}
  \includegraphics[width=0.33\linewidth]{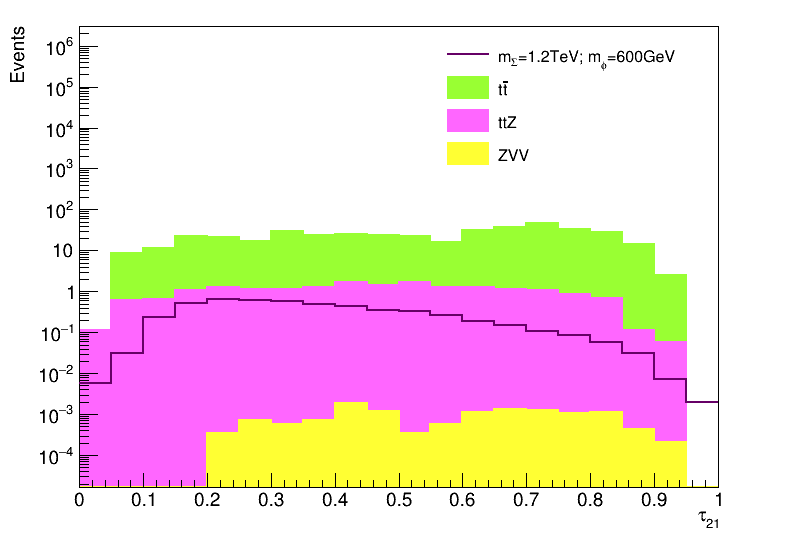}
  \includegraphics[width=0.33\linewidth]{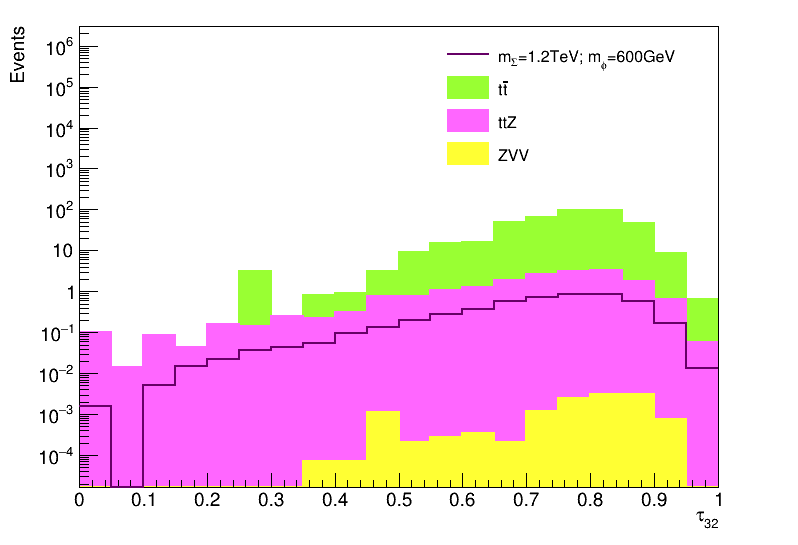}    
  \includegraphics[width=0.33\linewidth]{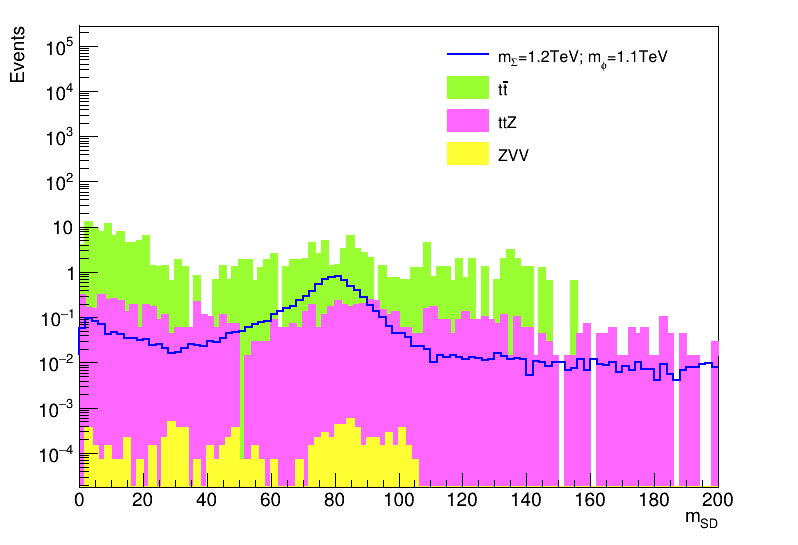}
  \includegraphics[width=0.33\linewidth]{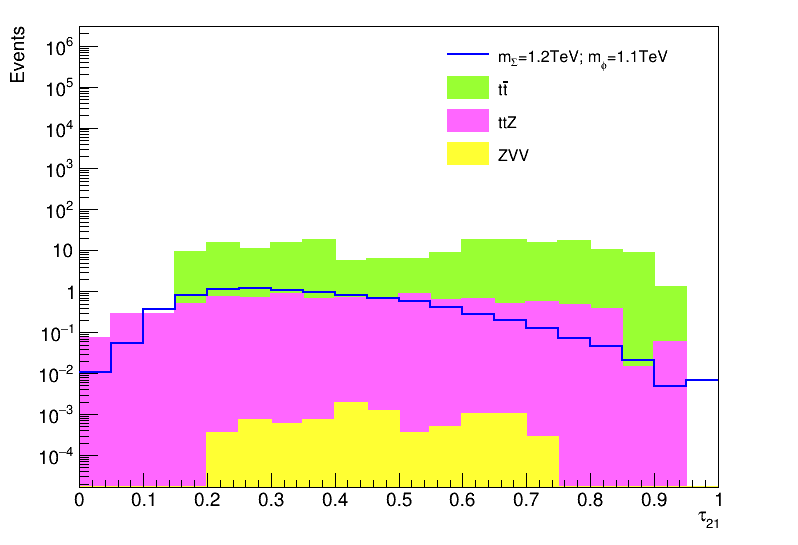}
  \includegraphics[width=0.33\linewidth]{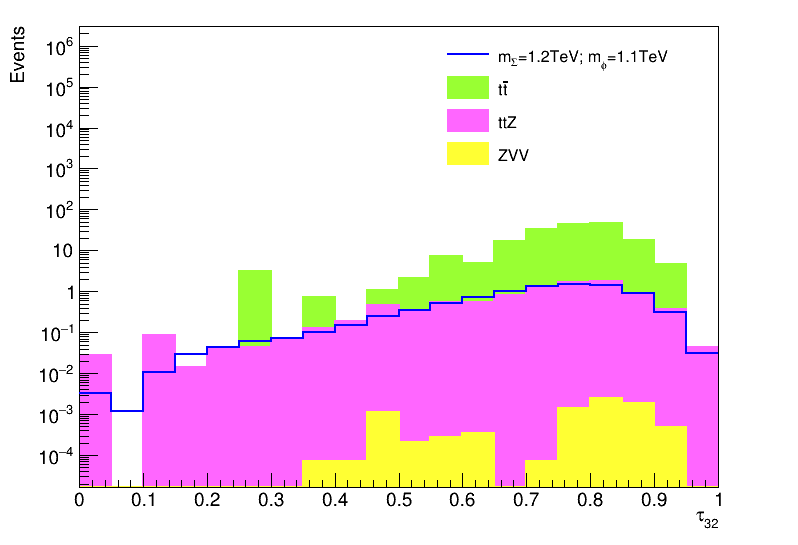}    
  \caption{The jet substructure variables for the low-mass (top), med-mass (middle), and high-mass (bottom) benchmark point
    and the background processes after the S2 selections are imposed in the $\fjtwo$ final state.}
\label{fig:dist4}
\end{figure}

The event selection is applied in three sequential stages: {\bf S1} is applied to all signal benchmark points, {\bf S2} is applied differently for the three benchmark
scenarios, and {\bf S3} is applied to the fatjets:
\begin{itemize}
\item S1: We require events to satisfy $\met>50\gev$, $m_{min}> 50\gev$, $\Ht>200\gev$, $\St>500\gev$, and $\Lt+\met>100\gev$. The distributions of some kinematic
  variables after passing S1 are shown in Fig.~\ref{fig:dist3}.
\item S2: We require $m_{min}>100\gev$, $dR_{min}>1.0$, and $0.7<JMF<1.1$ for all events. Additional selections are imposed on $\Ht,\Lt,\St$, and $\Lt+\met$, which are 
  summarized in Table~\ref{quint_2l2j_cut2}. The $\Lt+\met$ requirement is effective at suppressing the $t\bar{t}$ background
  for the low- and mid-mass points.
\item S3: We require the selected fatjets to satisfy $60< m_{SD}<100\gev$, $\tau_{21}<0.8$, $\tau_{32}>0.3$, and $\rho>-5.5$. The distributions of
  these variables after S2 are shown in Fig.~\ref{fig:dist4}.
  \end{itemize}
The cut-flow is shown in Table~\ref{quint_cutflow}.

\begin{table}[h]
  \centering
\begin{tabular}{|c|c|c|c|c|}
  \hline
  S2 & {\small $\Ht$ } & {\small $\Lt$ } & {\small $\Lt+\met$ } & {\small $\St$ }   \\
\hline
 low-mass & $>$500 & $>$100 & $>$300 & $>$1100 \\
 mid-mass & $>$500 & $>$100 & $>$500 & $>$1400 \\
 high-mass& $>$700 &$>$ 200 & $>$500 & $>$1400 \\
    \hline
  \end{tabular}
\caption{The selections implemented in S2 for the signal and background processes in the $\fjtwo$ final state. The selections are shown in units of \gev.}

  \label{quint_2l2j_cut2}
\end{table}

\subsection{$\fjthree$ Channel} 
\begin{figure}[h!]
\includegraphics[width=0.49\linewidth]{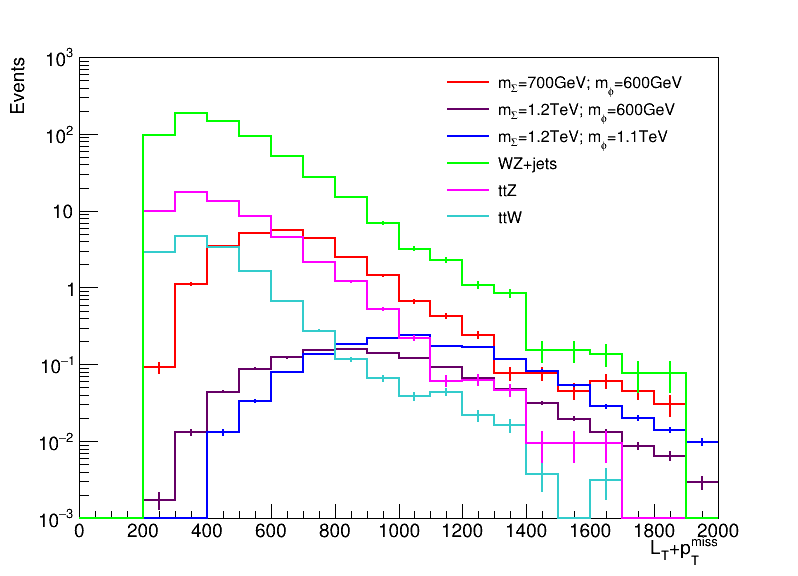}
\includegraphics[width=0.49\linewidth]{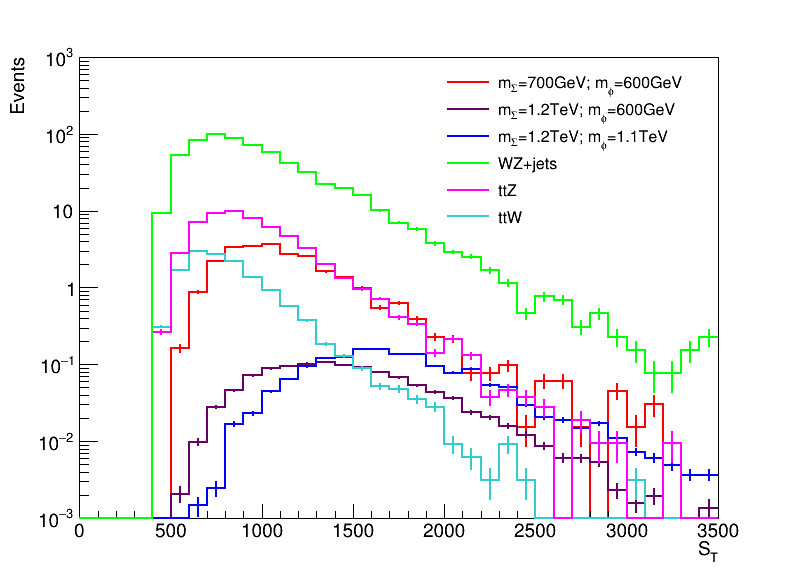}
\includegraphics[width=0.49\linewidth]{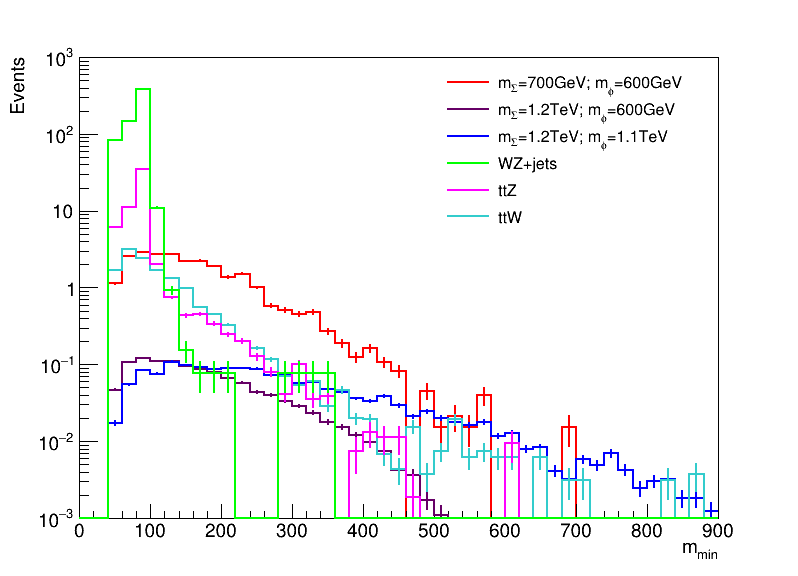}
\includegraphics[width=0.49\linewidth]{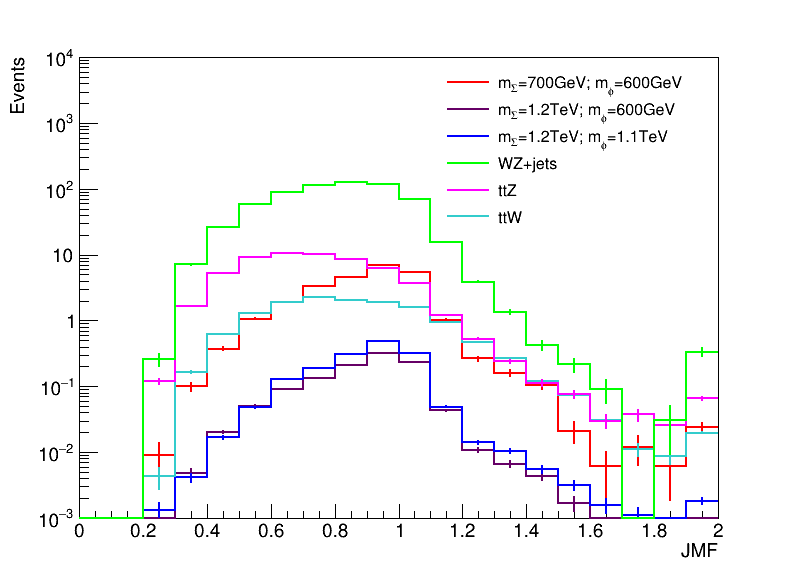}
\caption{The distributions of $\Lt+\met,\St$ (top row) and $m_{min}, JMF$ (bottom row) are shown for signal and background processes after
  the S1 selections are imposed in the \fjthree\ final state.}
\label{fig:dist5}
\end{figure}

\begin{figure}[h!]
  \includegraphics[width=0.33\linewidth]{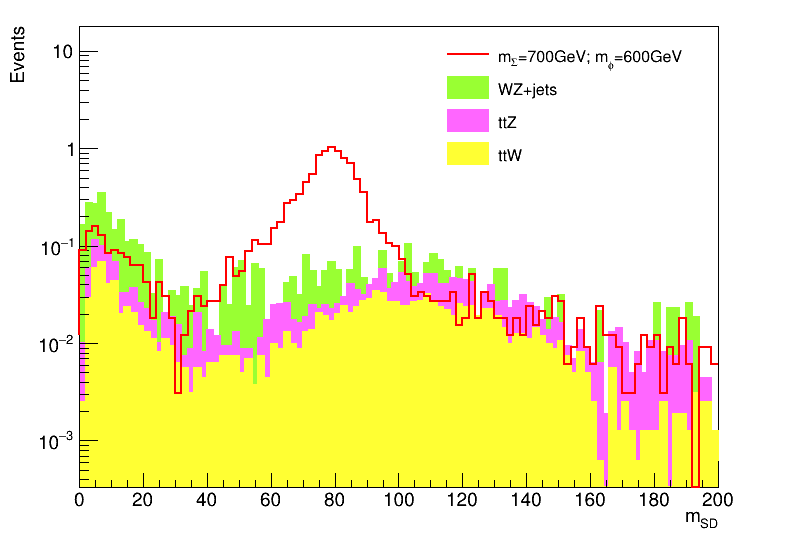}
  \includegraphics[width=0.33\linewidth]{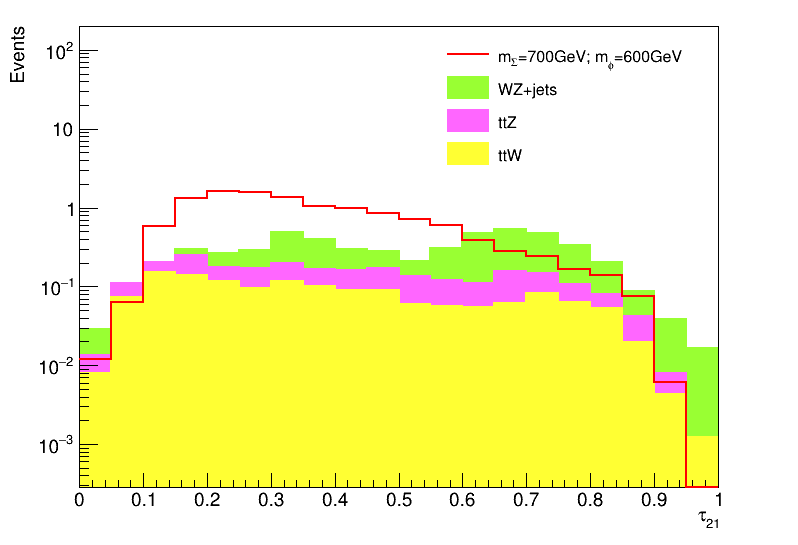}
  \includegraphics[width=0.33\linewidth]{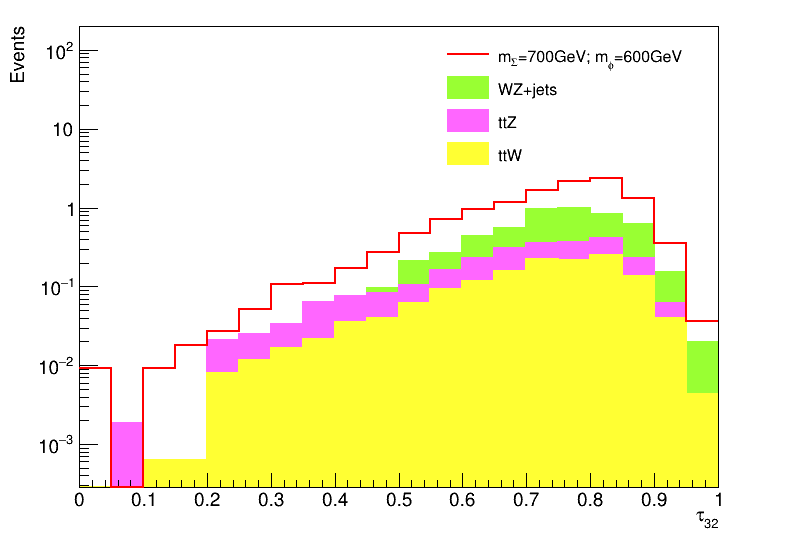}    
  \includegraphics[width=0.33\linewidth]{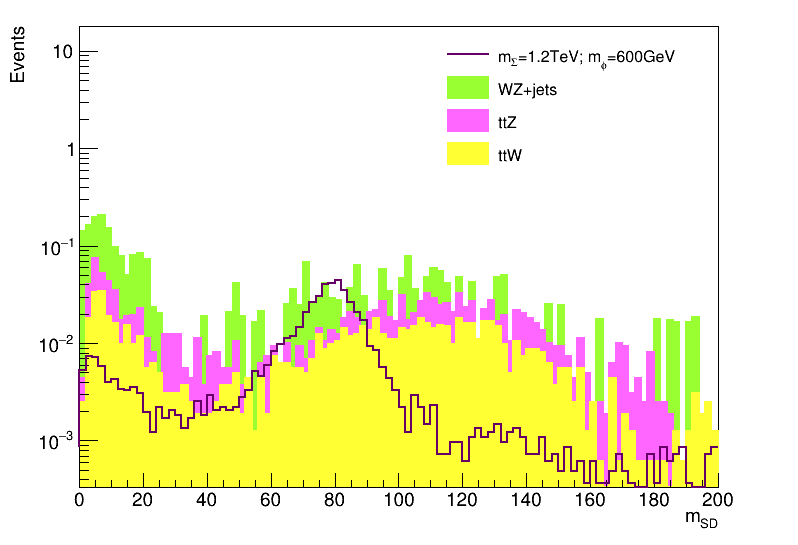}
  \includegraphics[width=0.33\linewidth]{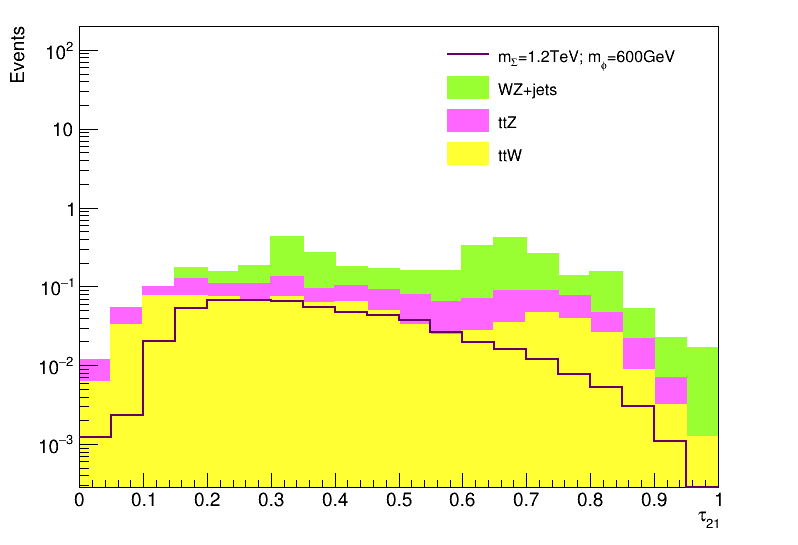}
  \includegraphics[width=0.33\linewidth]{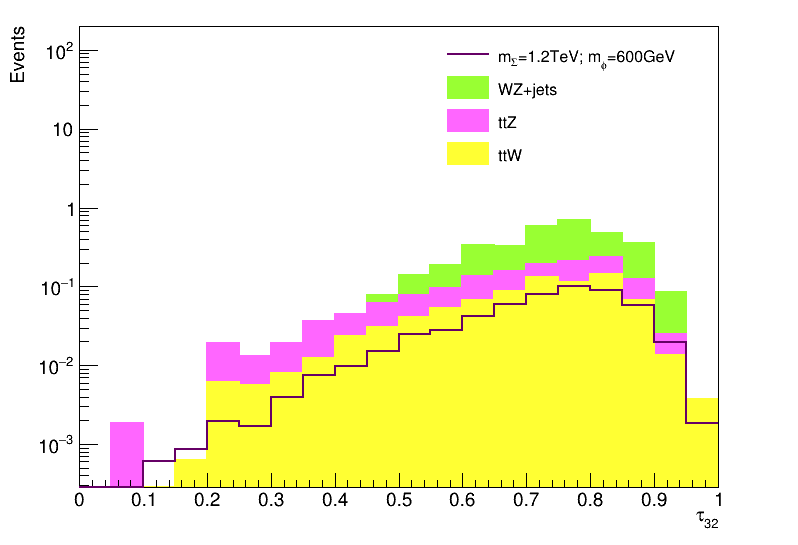}    
  \includegraphics[width=0.33\linewidth]{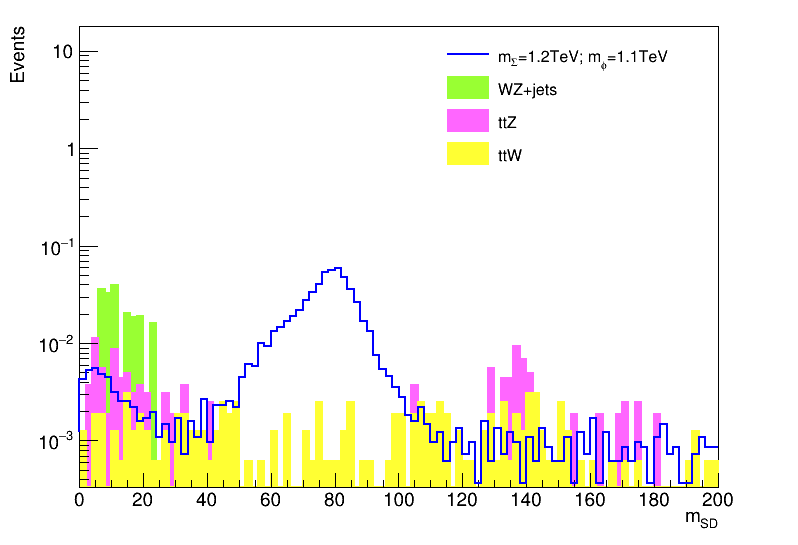}
  \includegraphics[width=0.33\linewidth]{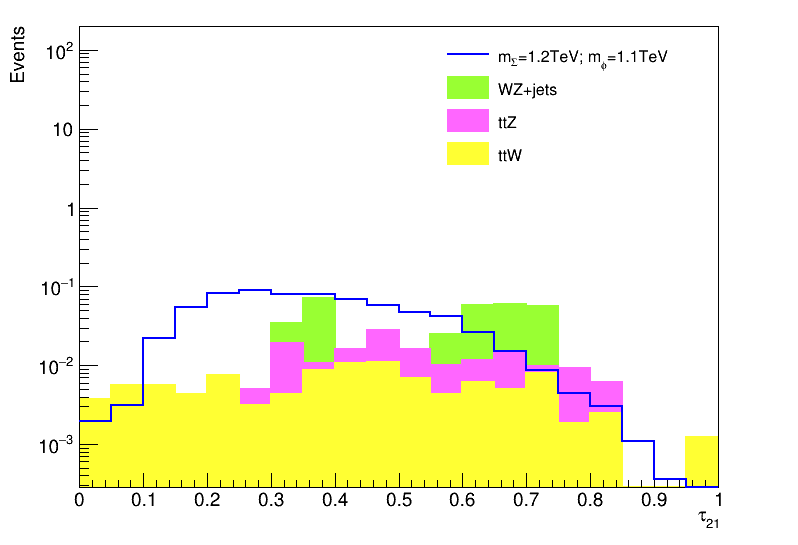}
  \includegraphics[width=0.33\linewidth]{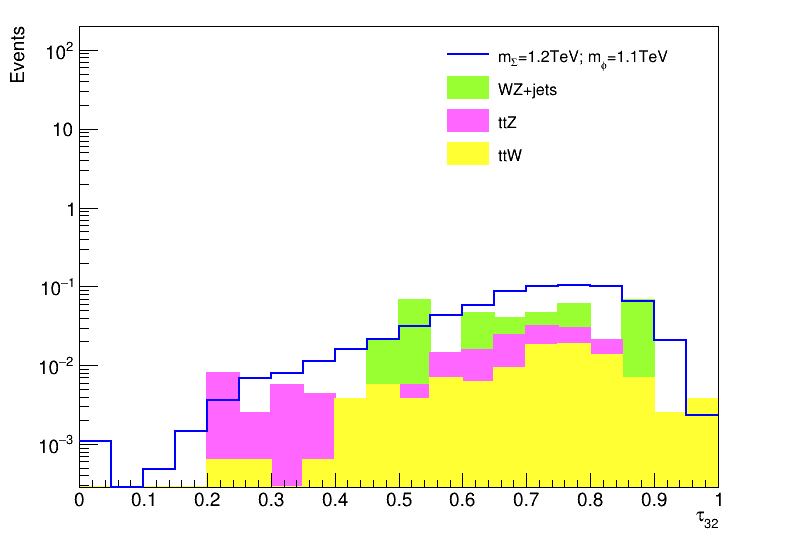}    
  \caption{The jet substructure variables for the low-mass (top), med-mass (middle), and high-mass (bottom) benchmark point
    and the background processes after the S2 selections are imposed in the $\fjthree$ final state.}
\label{fig:dist6}
\end{figure}

The \fjthree\ final state arises from the process $pp\rightarrow\Sigma^{++}\Sigma^{--}\rightarrow\phi^{++}\nu\phi^{--}\bar{\nu}\rightarrow W^+W^+\nu W^-W^-\bar{\nu}$
when one of the $W$ bosons decays hadronically, and all others decay leptonically. As in the case of \fjtwo, here as well we apply the event selection in
three sequential stages, S1, S2, and S3.
\begin{itemize}
  \item S1: We require events to satisfy $\met>50\gev$, $m_{min}> 50\gev$, $\Ht>200\gev$, $\St>400\gev$, and $\Lt+\met>200\gev$. The distributions of some kinematic
    variables after passing S1 are shown in Fig.~\ref{fig:dist5}. These follow a similar pattern as that in the \fjtwo\ final state. In this channel however,
    the smaller number of fatjets, and larger number of leptons requires changes to the $\Ht$ and $\Lt$ selections.
  \item S2: We require $m_{min}>100\gev$, $dR_{min}>1.0$, and $0.7<JMF<1.1$ for all events. Additional selections are imposed
    on $\Ht,\Lt,\St$, and $\Lt+\met$, which are   summarized in Table~\ref{quint_3l1j_cut2}.
    \item S3: We require the selected fatjet to satisfy $60< m_{SD}<100\gev$, $\tau_{21}<0.8$, $\tau_{32}>0.3$, and $\rho>-5.5$. The distributions of
      these variables after S2 are shown in Fig.~\ref{fig:dist6}.
\end{itemize}
The cut-flow is shown in Table~\ref{quint_cutflow}.

\begin{table}[h]
  \centering
\begin{tabular}{|c|c|c|c|c|}
  \hline
  S2 & {\small $\Ht$ } & {\small $\Lt$ } & {\small $\Lt+\met$ } & {\small $\St$ }   \\
\hline
 low-mass & $>$200 & $>$200 & $>$400 & $>$700 \\
 mid-mass & $>$200 & $>$200 & $>$500 & $>$800 \\
 high-mass& $>$400 & $>$400 & $>$700 & $>$1100 \\
    \hline
  \end{tabular}
\caption{The selections implemented in S2 for the signal and background processes in the $\fjthree$ final state. The selections are shown in units of \gev.}

  \label{quint_3l1j_cut2}
\end{table}

\subsection{Results}
We now summarize the results for the \fjtwo\ and \fjthree\ channels. Table~\ref{quint_cutflow} shows the cut-flow for the signal benchmark points after
S1, S2, and S3. After S2, we see that the background is comparable (or larger) than the signal in both the channels. However, implementing S3 reduces
the background significantly. The signal yield in the \fjthree\ channel is smaller than in the \fjtwo\ channel for the low-mass benchmark point.

\begin{table}[h!]
  \centering
  \renewcommand{\arraystretch}{1.5}
  \begin{tabular}{| p{1.2cm} | p{2.5cm} | p{1.5cm} | p{1.5cm} | p{1.5cm} | p{1.5cm} | p{1.5cm} | p{1.5cm} |}
    \hline
    \multirow{3}{1.2cm}{Final State} & \multirow{3}{2.9cm}{$m_{\Sigma^{++}},m_{\phi^{++}}$ (\GeV)} & \multicolumn{6}{|c|}{Cut Flow ($\times 10^{-3}$ fb)} \\
    \cline{3-8}
    & & \multicolumn{2}{|c|}{S1}& \multicolumn{2}{|c|}{S2} & \multicolumn{2}{|c|}{S3} \\
    \cline{3-8}
    & & Sig & Bkg & Sig & Bkg & Sig & Bkg \\
    \hline
    \multirow{3}{1.8cm}{$\fjtwo$} &  1200,1100 (high-mass)& \small{$47.1$} & \small{$51530$} & \small{$32.2$} & \small{$669.1$} & \small{$18.33$} & \small{$7.28$} \\
    \cline{2-8}
    &  1200,600 (med-mass) & \small{$28.1$} & \small{$51530$} & \small{$18.5$} & \small{$1506$} & \small{$9.36$} & \small{$9.74$} \\
    \cline{2-8}
    &  700,600 (low-mass) & \small{$613.8$} & \small{$51530$} & \small{$390.5$} & \small{$6592$} & \small{$208.50$} & \small{$182.60$} \\
    \hline
    \multirow{3}{1.8cm}{$\fjthree$} &   1200,1100 (high-mass) & \small{$11.5$} & \small{$5096$} & \small{$4.98$} & \small{$3.05$} & \small{$3.65$} & \small{$0.12$} \\
    \cline{2-8}
    &  1200,600 (med-mass) & \small{$8.2$} & \small{$5096$} & \small{$3.97$} & \small{$24.66$} & \small{$2.82$} & \small{$4.48$} \\
    \cline{2-8}
    &  700,600 (low-mass) & \small{$184.4$} & \small{$5096$} & \small{$86.90$} & \small{$39.01$} & \small{$65.20$} & \small{$7.33$} \\
    \hline

  \end{tabular}
  \caption{The cut flow for the \fjtwo\ and \fjthree\ final states for the different sets of selections applied.}
  \label{quint_cutflow}
\end{table}

The sensitivity of the signal is typically calculated with the formula: $Z_{\mathrm{simple}} = \frac{s}{\sqrt{b}}$ where $s$ is the number of signal events
and $b$ is the number of background events. This can overestimate the significance in cases where the background is very small. We instead
adopt the general formula~\cite{Cowan:2010js,Cousins:2007yta} to calculate the significance, which also permits incorporating the
uncertainty in the background measurements. The general formula is
\begin{equation}
Z_{\mathrm{general}} = \left[2\left(\left(s+b\right)\ln\left[\frac{\left(s+b\right)\left(b+\Delta_b^2\right)}{b^2+\left(s+b\right)\Delta_b^2}\right]-\frac{b^2}{\Delta_b^2}\ln\left[1+
\frac{s\Delta_b^2}{b\left(b+\Delta_b^2\right)}\right]\right)\right]^{1/2}.
  \label{sig}
\end{equation}
where, $\Delta_b$ is the uncertainty in the background estimation.
In Table~\ref{quint_sens} we have tabulated the significance using the general form with two choices of background uncertainty,
$\Delta_b=$ 10\% and 50\%. For comparison we also tabulate the significance $Z_{\mathrm{simple}}$. 
The significance is calculated at current and projected future luminosities of the LHC.
\begin{table}[h!]
  \centering
  \renewcommand{\arraystretch}{1.5}
  \begin{tabular}{| p{1.7cm} | p{2.1cm} | p{3.5cm} | p{3.5cm} | p{3.1cm} |}

    \hline
    \multirow{2}{1.7cm}{Final State} & \multirow{2}{2.9cm}{$m_{\Sigma^{++}},m_{\phi^{++}}$ (\GeV)} & \multirow{2}{3.5cm}{ $Z_{\mathrm{simple}}$} & \multicolumn{2}{|c|}{$Z_{\mathrm{general}}$} \\
    \cline{4-5}
    & & & $\Delta_b = 10\%$ & $\Delta_b = 50\%$ \\
    \hline
    \multirow{3}{1.8cm}{$\fjtwo$} & 1200,1100 (high-mass) & $2.53~(3.72)~[11.78]$ & $1.94~(2.84)~[7.85]$ & $1.65~(2.11)~[2.99]$ \\
    \cline{2-5}
    & 1200,600 (med-mass)& $1.12~(1.64)~[5.20]$ & $0.98~(1.42)~[3.912]$ & $0.82~(1.05)~[1.44]$ \\
    \cline{2-5}
    & 700,600 (low-mass) & $5.73~(8.45)~[26.72]$ & $4.30~(5.59)~[8.15]$ & $1.64~(1.70)~[1.74]$ \\
    \hline
    \multirow{3}{1.8cm}{$\fjthree$} & 1200,1100 (high-mass) & $3.90~(5.76)~[18.20]$ & $1.60~(2.36)~[7.41]$ & $1.59~(2.31)~[6.28]$ \\
    \cline{2-5}
    & 1200,600 (med-mass) & $0.50~(0.73)~[2.31]$ & $0.45~(0.66)~[1.96]$ & $0.42~(0.57)~[0.95]$ \\
    \cline{2-5}
    & 700,600 (low-mass) & $8.95~(13.19)~[41.72]$ & $5.21~(7.56)~[19.74]$ & $4.13~(5.14)~[6.91]$ \\
    \hline

  \end{tabular}
  \caption{The significance of \fjtwo\ and \fjthree\ channels calculated at integrated luminosities of 138, (300) and [3000]~\ifb including two
    estimates of uncertainties in background yield.}
  \label{quint_sens}
\end{table}

We can see from Table~\ref{quint_sens} that the low-mass benchmark point has the largest significance in both the channels. 
The significance in the low-mass point is more than $5\sigma$ at the current luminosity of the LHC experiments,
when no uncertainty in the background is assumed. However, with a background uncertainty $\Delta_b = 50\%$ , significance more than $5\sigma$
is achievable at $300~\ifb$ only in the \fjthree\ channel. The significance in the \fjtwo\ channel is comparatively less because the
signal and background yields are comparable. 
If we consider the high mass point, then the significance in both the channels is almost similar. However with $\Delta_b = 50\%$, 
the \fjthree\ channel offers a significance of more than $5\sigma$ with an integrated luminosity of  $3000~\ifb$.
Overall we found that the performance of both the channels is excellent in the low-mass and high-mass benchmark points and both of these channels
show potential for discovery (or stringent constraints) from current and future luminosities at the LHC experiments.

\section{\label{sec:4}Outlook}

To address the shortcomings of the standard model, extensions beyond the SM introduce new physics at or beyond the TeV scale, often including new particles.
Scenarios involving both fermion and scalar multiplets naturally arise in several BSM theories like composite Higgs models or grand unified theories. Such
scenarios can lead to unique collider signatures depending on the mass hierarchies between the new hypothetical particles. In this paper we focused on a
simplified model with a neutral fermion quintuplet and a scalar quadruplet, which can generate neutrino masses through tree and loop-level processes.
The quintuplet fermions decay into standard model gauge bosons via the scalars and produce distinctive signals at the LHC involving multilepton and
multijet final states. 

These final states can often have comparatively large SM backgrounds at the LHC given the low production cross sections of the new fermions, especially at around
1~\tev\ or higher fermion masses. This work examined the pair production of quintuplet fermions in the range (700 -1200)~\gev, whose decay eventually yields
highly boosted $W$ and $Z$ bosons. Here we focused on production of doubly charged fermions due to their higher cross sections over other pair production channels.
We used fatjets to target the boosted gauge boson decays, and applied advanced jet substructure techniques and dedicated variables to improve the LHC sensitivity in
two specific final states: \fjtwo\ and \fjthree.

We did detailed analysis of optimizing the signal selection against backgrounds for three benchmark scenarios with low, medium, and high masses of the
quintuplet fermions and scalars. A well chosen selection strategy ensured a high sensitivity to the signal over the background for the different mass scenarios.
We found that the low-mass scenario achieved the highest significance in both the \fjtwo\ and \fjthree\ channels, exceeding $5\sigma$ at current LHC luminosity
with no background uncertainty. With a conservative choice of $50\%$ background uncertainty, only the \fjthree\ channel reaches $\geq 5\sigma$ at $3000~\ifb$.
For the high-mass scenario, the significance is similar for the two channels with no background uncertainty. In the conservative case of $50\%$ background uncertainty,
here as well the \fjthree\ channel has $\geq 5\sigma$ significance with $3000~\ifb$.

Both channels maintain good sensitivity with a realistic background uncertainty, and will enable the LHC experiments to probe masses of more than $1\tev$ even with
the existing LHC dataset of $300~\ifb$ (Run 2 + Run 3), and higher masses with the projected future luminosities of HL-LHC.

\section*{Acknowledgments}
\vspace{-0.5cm}
N.K. would like to thank Anusandhan National Research Foundation
Government of India for the support from Core Research Grant with grant agreement number CRG/2022/004120. 
\vspace{-0.3cm}
\bibliographystyle{unsrt}
\bibliography{ref}
\end{document}